# Estimating public transport congestion in UK urban areas with open transport models

J. Raimbault[1*] and M. Batty[1]

[1]Centre for Advanced Spatial Analysis, University College London

February 15, 2021

**Summary**

Operational urban transport models require to gather heterogeneous sets of data and often integrate different sub-models. Their systematic validation and reproducible application therefore remains problematic. We propose in this contribution to build transport models from the bottom-up using scientific workflow systems with open-source components and data. These open models are aimed in particular at estimating congestion of public transport in all UK urban areas. This allows us building health indicators related to public transport density in the context of the COVID-19 crisis, and testing related policies.

## 1. Introduction

With the COVID-19 crisis and associated social distancing requirements, public transport has been designed and perceived as a risky environment (Tirachini and Cats, 2020), besides loosing ridership due to drops in mobility and modal shift (Gutiérrez et al., 2020). A detailed understanding of how work and transport related policies impacts effective densities in public transport is therefore an important issue. When systematically designing and testing transport policies, a reproducible and validated workflow is crucial first to ensure comparability between different case studies, but also to be able to apply them on a large set of urban areas.

Urban transportation models such as four-step models, and more generally land-use transport interaction models (Wegener, 2004), require the integration of heterogenous data and the coupling of various submodules with possibly high levels of complexity. This raises issues on the one hand for their implementation, transferability and reproducibility, and on the other hand for their validation which requires large scale numerical experiments to validate the submodules and the whole models (Lee, 1973; Batty, 2014). This work tackles both issues by leveraging modularity and transparency for the construction of large urban models in a modular way, using scientific workflow systems (Barker, 2007) to couple the different components of transport models and to launch numerical experiments for their validation.

We propose thus in this contribution to apply this modular model building methodology to construct open transport models and estimate public transport density. The approach is systematically applied on all UK urban areas, and can be leveraged to investigate policies to mitigate propagation related to mobility and public transport.

## 2. Methods

We implement this approach by building a modular four-step multimodal transportation model using only open-source projects. We couple together the MATSim model (MATSim Community (Horni et al., 2016)) to simulate the transportation system, the SPENSER model (University of Leeds, https://github.com/nismod/microsimulation) for the generation of synthetic population, the QUANT model (University College London (Milton and Roumpani, 2019)) to estimate spatial interactions, and the spatialdata library (OpenMOLE Community (Raimbault et al., 2020)) for data preparation.

---

1       * j.raimbault@ucl.ac.uk

Data used to construct the synthetic population is mostly Census data, while transportation networks are built combining Ordnance Survey open data, OpenStreetMap data, and GTFS data for public transport timetables. The QUANT model uses census commuting flows to estimate spatial interaction parameters.

The model components are embedded as docker containers into the DAFNI facility (https://dafni.ac.uk/), which workflow system is used to couple them and build the integrated model. DAFNI provides a scientific workflow system for model integration and coupling, direct access to relevant open datasets, visualisation functionalities, and access to a High Performance Computing infrastructure. We show in Figure 1 the workflow constructed with the interactive workflow builder within the platform.

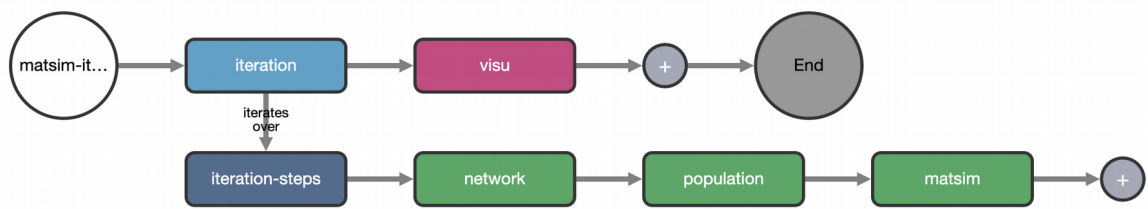

**Figure 1** Construction of the transport model using the DAFNI workflow system: DAFNI workflow including model steps and a computational experiment (Monte Carlo simulations)

## 3. Results

The model is run on the largest functional urban areas (following the definition of (Florczyk et al., 2019)) in the UK. We show first results of numerical experiments studying the role of stochasticity on model outputs, for example in Figure 2 for the statistical distribution of trip departure times (these are iteratively evolved by agents in the MATSim model) for the urban area of Taunton. We also show in Figure 2 the distribution of car trip distances in the different urban areas, for 13 large urban areas (excluding London for performance purposes).

Source code to prepare the model components, input data, and docker containers is available on an open-source git repository at https://github.com/JusteRaimbault/UrbanDynamics. To illustrate the reproducibility of our approach, we test the construction of the model with the OpenMOLE workflow engine (Reuillon, 2013), which provides a scripted workflow engine and methods to calibrate and validate simulation models, and suggest advanced numerical experiments for the validation of the coupled model. For example, studying the role of spatial configuration on model outcomes (Raimbault et al., 2019) would be relevant to understand the influence of missing or imprecise data and sampling for the synthetic population. Furthermore, the implementation as open source OpenMOLE scripts embedding open models is an asset for an easier teaching of transport models, as these can seamlessly be run and explored by students without the need to access particular resources or proprietary software (Carrington & Kim, 2003).

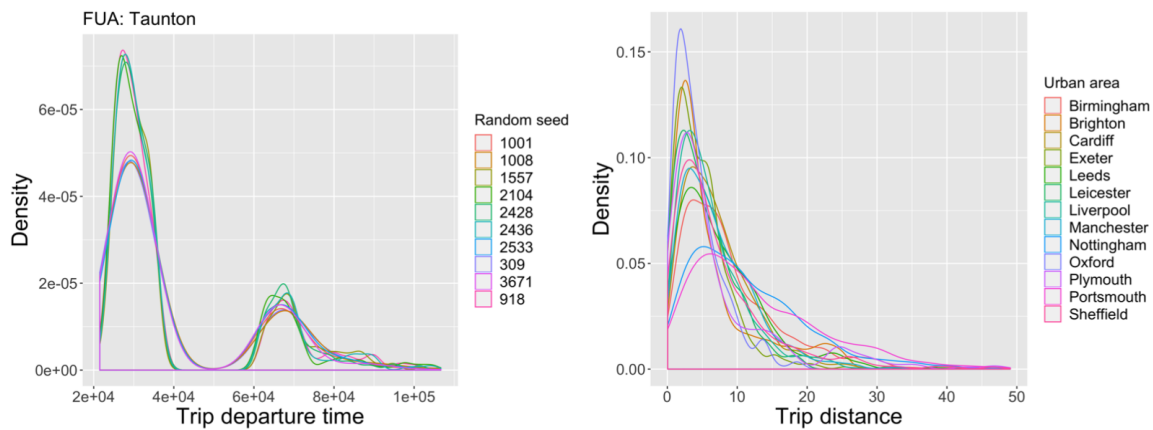

**Figure 2** Results of the simulation of the integrated model on the largest functional urban areas in the UK. (Left) Distribution of trip departure times, for several stochastic repetitions on the same urban area; (Right) Distribution of trip distances for all urban areas (London was not included yet for performance reasons as numerical experiments are still being streamlined).

### 4. Towards health indicators for public transport

Work in progress includes the application of this model to the development of health indicators within public transportation, and more particularly linking transportation and work-from-home policies with effective densities in public transport which provide potential exposure indicators in the context of the COVID-19 crisis. While urban density in itself is not significantly linked to stronger transmission (Hamidi et al., 2020), possible transmission within closed environment of public transport causes some concerns. The MATSim transport model has already been applied in the pandemic context: for example, Müller et al. (2020) build on the MATSim model to construct a full epidemiological agent-based model for the propagation of COVID-19 for the Berlin urban area. We limit our work to estimating public transport congestion and density, providing a measure of potential contacts between individuals. Our approach is therefore easier to test and validate, as less sub-models and aspects have to be tested, and we expect it to provide more robust conclusions on public transport itself. As our transport model can be systematically applied to any urban area in UK, we also expect to be able to test and compare policies across different urban and epidemiological contexts.

### 5. Acknowledgements

The authors acknowledge the DAFNI platform for funding through the Champions program, and the funding of Urban Dynamics Lab Grant EPSRC EP/M023583/1.

**Biographies**

**Juste Raimbault** is a Research Fellow at the Centre for Advanced Spatial Analysis, University College London. His research interests include the modeling and simulation of complex urban systems, land-use transport interactions, agent-based modeling, artificial life, and bibliometrics. He is a contributor to the OpenMOLE platform for model exploration and validation.

**Michael Batty (CBE FRS FBA)** is Bartlett Professor of Planning at University College London where he is Chair of the Centre for Advanced Spatial Analysis (CASA). He has worked on computer models of cities and their visualisation since the 1970s and has published many books. He is one of the most highly cited academics in city planning and human geography with an H index of 95 and 39000 citations.